\documentclass[secnumarabic,aps,prl,reprint]{revtex4-2}
\usepackage[english]{babel}
\usepackage{graphicx}
\usepackage{color}
\usepackage{amsmath}
\usepackage{tabularx}
\usepackage[dvipsnames]{xcolor}

\begin{document}

	\title{Symmetry constraints on the orbital transport in solids}
	\author{Sergei Urazhdin}
	\affiliation{Department of Physics, Emory University, Atlanta, GA, USA.}

\begin{abstract}
We show that electron interaction with the crystal lattice imposes stringent symmetry constrains on the atomic orbital moment propagation. We present examples that elucidate the underlying  mechanisms and reveal an additional effect of ultrafast orbital moment oscillations not captured by the semiclassical models. The constraints revealed by our analysis warrant re-interpretation of prior observations, and suggest routes for efficient orbitronic device implementation.
\end{abstract}

\maketitle

\begin{figure}
	\centering
	\includegraphics[width=\columnwidth]{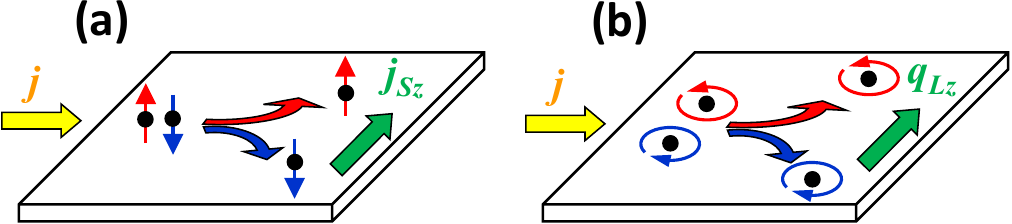}
	\caption{\label{fig:Hall} Analogy between SHE (a) and OHE (b).}
\end{figure}

Spin-electronic phenomena such as spin transfer torque (STT)~\cite{Slonczewski1996}, Rashba-Edelstein and spin Hall effect (SHE)~\cite{Dyakonov1971,Rashba2005} have attracted intense interest as important manifestations of the interplay between electron's spin and orbital degrees of freedom, with applications in sensing and information technology. SHE, one of the most promising mechanisms for spin current generation, is facilitated by spin-orbit coupling (SOC), which couples spin to chiral orbital transport~\cite{RevModPhys.87.1213}. Bypassing the requirement for a large SOC, and directly using orbital moments may enable efficient low-cost orbitronic devices based on light elements instead of the heavy metals in SOC-based devices~\cite{jungwirth2012spin}. This possibility has motivated a flurry of research into orbital moment generation and transport effects analogous to those involving spin, including orbital Hall effect (OHE) as a counterpart to SHE enabling efficient generation of orbital moment in light transition metals~\cite{PhysRevLett.100.096601,PhysRevB.77.165117,PhysRevLett.102.016601,PhysRevB.98.214405,PhysRevB.106.104414} illustrated in Fig.~\ref{fig:Hall}, orbital torque that similarly to STT may enable control of magnetic moments~\cite{Mokrousov2020,Lee2021-sf,PhysRevResearch.4.033037}, and long-range orbital moment transport~\cite{Go2021-sn,PhysRevB.107.134423,Hayashi2023-rg}. However, the differences between spin and orbital counterparts remain largely unexplored.

Spin carried by the Bloch waves is only weakly perturbed by the lattice via SOC. In contrast, we show below that atomic orbital moment dynamics can be dominated by interaction with the lattice potential even in the absence of SOC. The resulting effects are dependent on the orientation of the orbital moments and the propagation direction, imposing hitherto largely unrecognized symmetry constraints on orbitronic device geometries and structure qualitatively distinct from spintronics.

The concept of transport of physical quantities emerges from the relations connecting their local variations to flux. For the density operator $\hat{l}_\alpha(\mathbf{r},t)$ of $\alpha^{th}$ component of orbital angular momentum $\hat{\mathbf{L}}=\mathbf{r}\times\hat{\mathbf{p}}$ of an electron moving in the potential $U(\mathbf{r})$, this relation is~\cite{PhysRevB.77.165117,Mokrousov2020}
\begin{equation}\label{eq:continuity}
	\frac{\partial \hat{l}_\alpha(\mathbf{r},t)}{\partial t}=-\mathbf{\nabla}\cdot\hat{\mathbf{q}}_{L_\alpha}+\hat{T}_\alpha.
\end{equation}  
Here, $\hat{\mathbf{q}}_{L_\alpha}$ is the density of the orbital angular momentum current
\begin{equation}\label{eq:flux}
\hat{\mathbf{Q}}_{L_\alpha}=\{\hat{L}_\alpha,\hat{\mathbf{v}}\}/2,
\end{equation}
where $\hat{\mathbf{v}}$ is the electron velocity operator, and 
$\hat{T}_\alpha$ is the density of the $\alpha^{th}$ component of orbital torque 
\begin{equation}\label{eq:torque}
	\hat{\mathbf{\Gamma}}=\frac{i}{\hbar}[U,\hat{\mathbf{L}}]=\mathbf{r}\times\mathbf{F}(\mathbf{r})
\end{equation}  
exerted by $U(\mathbf{r})$, where $\mathbf{F}=-\mathbf{\nabla}U$ is the force.

According to Eq.~(\ref{eq:continuity}), the rate of variation of angular momentum in a unit volume is determined by its flux through the boundary plus the torque exerted by $U$. In the absence of torque, it reduces to the continuity relation 
\begin{equation}\label{eq:continuity2}
	\frac{\partial \hat{l}_\alpha(\mathbf{r},t)}{\partial t}=-\mathbf{\nabla}\cdot\hat{\mathbf{q}}_{L_\alpha},
\end{equation}
which signifies the conservation of orbital angular momentum, enabling control of the local orbital moment in a certain volume by its injection into this volume, for example via OHE. However, if the torque dominates the dynamics described by Eq.~(\ref{eq:continuity}), orbital current produced by OHE cannot be used to control orbital moments. In particular, accumulation of orbital moment generated by OHE can be prevented by its precession due to the crystal-field torque analogous to spin precession due to the Larmor torque produced by the magnetic field. 

To elucidate the effects of orbital torque, consider the z-component of orbital angular momentum $\hat{L}_z=x\hat{p_y}-y\hat{p}_x=-i\hbar\partial/\partial\phi$, where $\phi$ is the polar angle in cylindrical coordinates. The corresponding component of torque is
\begin{equation}\label{eq:Tz}
	\hat{\Gamma}_z=y\frac{\partial U}{\partial x}-x\frac{\partial U}{\partial y}=-\frac{\partial U}{\partial \phi}.
\end{equation}
As expected from the Noether's theorem, the continuity relation Eq.~(\ref{eq:continuity2}) is satisfied if the potential is symmetric with respect to rotation around the z-axis.

In crystals, continuous rotational symmetry is broken by the lattice. Nevertheless, if the Fermi surface is well-described by the free electron approximation, the effects of lattice potential are small so the continuity equation Eq.~(\ref{eq:continuity2}) is expected to provide a good approximation for orbital dynamics. This can be seen from Eq.~(\ref{eq:Tz}). If the wavefunction does not significantly vary over the lattice constant, one can replace $\frac{\partial U}{\partial \phi}$ with its integral over a sphere centered at $r=0$, which identically vanishes. 

In orbitronic devices based on transition metals~\cite{PhysRevB.77.165117,PhysRevLett.102.016601,PhysRevB.98.214405,PhysRevResearch.4.033037}, most of the angular momentum is expected to be carried by the atomic moments of d-electrons, for which such an averaging procedure is unjustified since the wavefunction phase varies over at least $2\pi$ within a single atomic site. Our main result is the prediction of a large non-classical orbital crystal torque which can suppress certain components of propagating orbital moment over essentially a single lattice constant. The identified mechanism does not contradict the possibility of local orbital moment generation via OHE or its importance as the mechanism underlying SHE~\cite{PhysRevLett.121.086602}. Nevertheless, it places stringent constraints on the possibility of atomic orbital moment diffusion and accumulation over distances exceeding a few lattice constants~\cite{PhysRevLett.121.086602,Mokrousov2020}. 

First, we consider the conduction band of complex transition metal oxides exemplified by SrTiO$_3$, which is dominated by the $t_{2g}$ orbitals of the transition metal atoms on a cubic lattice~\cite{Tokura2000-tc,Dylla2019}.  Each of the $t_{2g}$ orbitals hybridizes via oxygen atoms located on the lines between transition metal atoms with only four of the six nearest neighbors, e.g. the $d_{xy}$ orbital hybridizes with four $d_{xy}$ orbitals of the nearest neighbors in the xy plane~\cite{Dylla2019,Urazhdin_STO}. The corresponding  tight-binding Hamiltonian is
\begin{equation}\label{eq:H_hop_r}
	\hat{H}=-V\sum_{\mathbf{n},\mathbf{l},m,s} (1-\delta_{l,m})\hat{c}^+_{\mathbf{n}+\mathbf{l},m,s}\hat{c}_{\mathbf{n},m,s},
\end{equation}
where $d_m=(d_{yz},d_{xz},d_{xy})$, the operator $\hat{c}^+_{\mathbf{n},m,s}$ creates an electron on site $\mathbf{n}$ with spin $s$ in the orbital state $m$, $\mathbf{l}$ is a unit vector in one of the six principal directions, and $V$ is the hopping matrix element describing orbital-selective hybridization. In the reciprocal space,
\begin{equation}\label{eq:H_hop_k}
	\hat{H}=\sum_{\mathbf{k},m,s}\epsilon_m(\mathbf{k})\hat{c}^+_{\mathbf{k},m,s}\hat{c}_{\mathbf{k},m,s},
\end{equation}
where $\hat{c}^+_{\mathbf{k},m,s}=\frac{1}{\sqrt{N}}\sum_{\mathbf{n}} e^{ia\mathbf{k}\mathbf{n}}\hat{c}^+_{\mathbf{n},m,s}$, $N$ is the number of lattice sites, $a$ is the lattice constant, with dispersion
\begin{equation}\label{eq:dispersion_STO}
\epsilon_m(\mathbf{k})=-2V\sum_{m'\ne m}\cos(k_{m'}a).
\end{equation}
This spectrum is orbitally degenerate along the planes $k_i=k_j$, which allows Bloch states to carry angular momentum~\cite{PhysRevLett.130.246701}. Consider, for example, a superposition of the Bloch states formed by orbitals $d_{xz}$ and $d_{yz}$, 
\begin{equation}\label{eq:Psi_STO}
\psi_{\mathbf{k},\sigma,s}=\frac{1}{\sqrt{2}}(\hat{c}^+_{\mathbf{k},xz,s}+i\sigma\hat{c}^+_{\mathbf{k},yz,s})|0\rangle,
\end{equation}
where $\sigma=\pm1$. They are stationary states (Bloch waves) in the two planes $k_x=\pm k_y$ at the intersection between the $d_{yz}$ and $d_{xz}$ sub-bands.

The contribution to orbital moment from the crystalline momentum identically vanishes for the plane wave by symmetry. On the other hand, the z-component of the atomic orbital momentum carried by this wave is 
$$
\langle\hat{L_z}\rangle=\langle\psi_{\sigma,\mathbf{k},s}|\sum_{\mathbf{n}}\hat{l}_z(\mathbf{n})|\psi_{\sigma,\mathbf{k},s}\rangle=\sigma\hbar,
$$
where $\hat{l}_z(\mathbf{n})$ is the z-component of atomic angular momentum on site $\mathbf{n}$,
and we used $\hat{l}_zd_{xz}=id_{yz}$, $\hat{l}_zd_{yz}=-id_{xz}$. 

Because of the anisotropy of the subbands $\epsilon_{xz}(\mathbf{k})$, $\epsilon_{yz}(\mathbf{k})$, the component $d_{xz}$ of the wave cannot propagate in the $y$-direction, while the component $d_{yz}$ - in the $x$-direction. Thus, the dispersion of the wavepackets formed by the states Eq.~(\ref{eq:Psi_STO}) is minimized for wavevectors along the $z$-axis. We conclude that orbital momentum along one of the principal axis is conserved by electrons propagating along this axis. The conservation of orbital angular momentum is ensured by the orbital selectivity of hopping, making complex oxides attractive for orbitronic applications. The requirement $\mathbf{L}\parallel\mathbf{k}$ for orbital moment conservation has been identified for other materials~\cite{PhysRevLett.130.246701}, and will be shown in another example below, suggesting its general importance.

For $k_x\ne\pm k_y$, $\psi_{\sigma,\mathbf{k},s}$ is not a stationary state, resulting in the evolution of the relative phase between its $d_{xz}$ and $d_{yz}$ components according to
\begin{equation}\label{eq:Psi_STO_t}
	\psi_{\sigma,\mathbf{k},s}(t)=\frac{1}{\sqrt{2}}(\hat{c}^+_{\mathbf{k},2,s}+ie^{it(\epsilon_2(\mathbf{k})-\epsilon_1(\mathbf{k}))/\hbar}\sigma\hat{c}^+_{\mathbf{k},1,s})|0\rangle.
\end{equation}
This state is characterized by the oscillating angular momentum 
$$
\langle\hat{L}_z\rangle=\sigma\hbar\cos[t(\epsilon_2(\mathbf{k})-\epsilon_1(\mathbf{k}))/\hbar].
$$
The flux divergence vanishes, so this oscillation cannot be described by the continuity relation Eq.~(\ref{eq:continuity2}). It results entirely from the torque exerted by the crystal potential described by the orbitally-selective Hamiltonian. The expectation values of both $\hat{L}_x$ and $\hat{L}_y$ remain zero, so the oscillation cannot be interpreted as precession of semi-classical angular momentum vector. Using $V=0.2$~eV, we estimate that the frequency of oscillation ranges from zero at the center of the Brillouin zone (BZ) to $10^{14}$~Hz at the BZ boundary along the $k_x$ or $k_y$ axes.

The underlying mechanism is similar to STT in ferromagnets~\cite{Slonczewski1996,Ralph2007SpinTT}. In STT, an electron is injected into a ferromagnet with its spin polarization non-collinear with the magnetization. Since the band structure is split into spin-dependent sub-bands, this state is not an eigenstate of the Hamiltonian, resulting in oscillation of the relative phase between the spin-up and spin-down components of the wavefunction manifested as spin precession. This dynamics involves angular momentum exchange between the magnetization and the injected spin producing STT. In the considered example of orbital dynamics, the role of exchange torque is played by the orbital torque produced by the crystal potential, and the reciprocal effect of this torque is a periodic rotation of the lattice. Semiclassical precession of orbital moment is not possible in this case, because angular momentum operator does not have matrix elements between $\psi_{+,\mathbf{k},s}$ and $\psi_{-,\mathbf{k},s}$.

Since orbital dynamics depends on the wavevector, oscillation of the orbital moment carried by a wavepacket decays due to dephasing between different momentum components, similar to spin dephasing in STT~\cite{Slonczewski1996,Ralph2007SpinTT}. Consider a Gaussian wavepacket centered around $\mathbf{k}_0\parallel\hat{x}$ and wavevector spread $\Delta k$. The orbital moment dephases over the time interval 
$$
\Delta t=\frac{\hbar}{\Delta k\frac{\partial\epsilon_2}{\partial k_x}|_{\mathbf{k}=\mathbf{k}_0}}=\frac{1}{\Delta k v_g},
$$
where $v_g$ is the group velocity. Since the wavepacket width is $\Delta x=\Delta k^{-1}=v_g\Delta t$, the components of orbital moment orthogonal to the direction of propagation decay over the packet width.

\begin{figure}
	\centering
	\includegraphics[width=0.9\columnwidth]{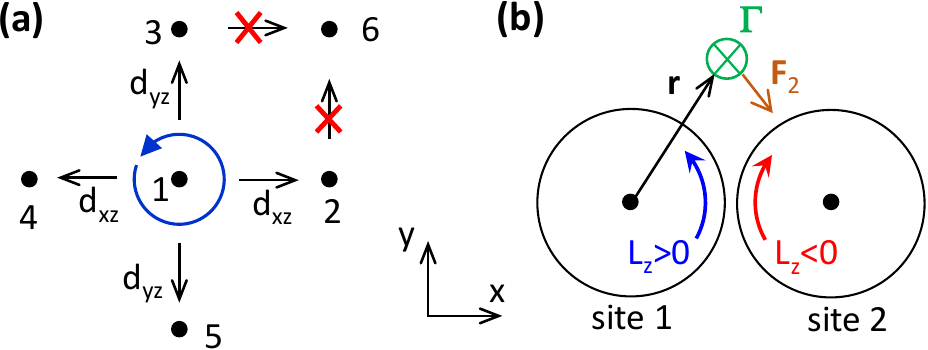}
	\caption{\label{fig:STO} (a) Orbitally-selective hopping of electron initially localized on site 1 in a cubic complex oxide. (b) Mechanism of orbital selectivity of hopping in the direction transverse to the orbital moment.}
\end{figure}

To elucidate the underlying microscopic mechanism, we consider an electron with orbital moment $L_z=\hbar$ initially localized on site 1, Fig.~\ref{fig:STO}(a). Based on the above analysis, orbital moment is expected to decay as the wavefunction spreads over the wave packet width, in this case a single lattice spacing. Indeed, because of the orbital selectivity of hopping, orbital component $d_{xz}$ flows to sites $2$ and $4$ along the x-axis, while orbital component $d_{yz}$ flows to sites $3$ and $5$ along the y-axis. The same orbital selectivity prevents reconstruction of the finite-$L_z$ orbital state by mixing between $d_{xz}$ and $d_{yz}$ along the diagonal, as shown in Fig.~\ref{fig:STO}(a) for site 6. Thus, a single electron hop along one of the principal axes quenches orbital moment normal to this axis, on the time scale $\Delta t\sim \hbar/V\sim 10^{-15}$~s. The possibility to transport $L_z$ in the xy plane by a wave packet with $k_x=\pm k_y$ is illusory, since the components $d_{xz}$ and $d_{yz}$ of the wave packet propagate in orthogonal directions along the $x$- and $y$-axes, respectively.

We now demonstrate similar symmetry constraints on orbital moment dynamics for a system with a completely different symmetry, a triangular 2d lattice which approximates a single $\{111\}$ fcc plane or $(0001)$ hcp plane in transition metals studied in the context of OHE~\cite{PhysRevLett.102.016601,PhysRevB.98.214405,Lee2021-sf,PhysRevResearch.4.033037,PhysRevB.107.134423,Hayashi2023-rg}. The hexagonal symmetry of the crystal field of triangular lattice does not quench the normal component of angular momentum of d-electrons, providing a close approximation for the axial rotational symmetry which according to Eq.~(\ref{eq:Tz}) allows its conservation. 

To the best of our knowledge, the d-orbital composition of the Bloch states on the triangular lattice does not have a simple form amenable to the analysis of orbital dynamics, due to the mismatch between d-orbital symmetries and the lattice geometry~\cite{Urazhdin_F}. We avoid this issue by considering an electron initially localized on a single site with the projection $L_z$ of orbital moment on the z-axis normal to the plane. We choose $L_z=2\hbar$ for concreteness, but the results are similar for other values of $L_z\ne0$. Orbital moment evolution is determined by hopping to the six nearest neighbors. By symmetry, hopping amplitudes onto the same orbitals of different neighbors are the same aside from the phase, so it is sufficient to consider a single neighbor on the x-axis, Fig.~\ref{fig:STO}(b). In the Slater-Koster method, hopping between neighboring d-orbitals is described by the parameters 
$V_{dd\sigma}$, $V_{dd\pi}=-2V_{dd\sigma}/3$, and $V_{dd\delta}=V_{dd\sigma}/6$~\cite{Harrison}. Their specific values, which can be calculated in the muffin-tin approximation, are not important. The matrix elements $V_{L_zL'_z}$ (with indices scaled by $\hbar$) describing hopping from $L_z=2\hbar$ onto the orbitals $L'_z$ are $V_{22}=0.06V_{dd\sigma}$, $V_{21}=V_{2-1}=0$, $V_{20}=-0.36V_{dd\sigma}$, $V_{2-2}=0.73V_{dd\sigma}$.
The probability that orbital moment is conserved on hopping is about $30$ times smaller than the probability that it is lost ($L'_z=0$), and about $150$ times smaller than the probability that it is reversed. 

This somewhat counter-intuitive orbital selectivity favoring orbital moment reversal results from the constructive interference between opposite-moment orbitals which remain in phase over the region of their overlap illustrated by curved arrows in Fig.~\ref{fig:STO}(b), as opposed to partially destructive interference of same-moment orbitals. Qualitatively, electron initially moving counterclockwise ($L_z>0$) around site 1 in the region between the sites, is moving clockwise with respect to site 2. As a consequence, there is a finite probability that it continues its motion as clockwise rotation around site 2, i.e. hopping tends to flip orbital moment normal to the hopping direction. The vanishing amplitudes $V_{2\pm1}$ ensure that orbital moment remains normal to the plane, i.e. the evolution is non-classical as in the previous example.

To put this dynamics in the framework of the continuity relation Eq.~(\ref{eq:continuity}), consider a volume surrounding site 1 and its nearest neighbors, such that at sufficiently short times the flux through its surface is negligible. The relaxation of orbital moment is caused by the crystal orbital torque exerted on the electron as it hops from site 1 to its nearest neighbors and is this directly related to the hopping mechanism itself, as illustrated in Fig.~\ref{fig:STO}(b).

We conclude that hopping in the direction normal to the orbital moment results in its relaxation over a single  lattice constant. On the other hand, similar analysis for the orbital moment initially aligned with the x-axis shows that orbital moment along the hopping direction is conserved, as in the previous example. Thus, despite substantial differences between the two considered systems, both exhibit the same symmetry constraints on the atomic orbital moment dynamics in relation to hopping direction. In contrast to cubic complex oxides, for the hexagonal 2d lattice crystal torque-mediated orbital relaxation occurs for any in-plane moment direction, because hopping to at least some of the six nearest neighbors is always non-collinear with the orbital moment. By the same argument, relaxation within distances comparable to the lattice constant is expected for any orbital moment direction on the 3d fcc or hcp lattice. This conclusion is consistent with numeric calculations, which show that orbital moment accumulation due to OHE in transition metals is limited to a few atomic constants from their surfaces or interfaces~\cite{PhysRevLett.121.086602,Mokrousov2020,PhysRevB.103.L121113}.

In summary, we analyzed two systems whose highly symmetric crystal fields allow unquenched atomic orbital moments. Our analysis reveals a dramatic anisotropy of atomic angular momentum dynamics dependent on its direction relative to  electron transport and crystal axes, placing stringent symmetry constraints on the possible structure and geometry of orbitronic devices. Orbitally-selective electron hopping along principal axes in cubic transition metal oxides such as SrTiO$_3$ preserves the component of atomic orbital moment in the hopping direction, making such materials particularly attractive for orbitronic applications. On the other hand, the normal component of atomic orbital moment is suppressed over essentially a single lattice spacing, making it impossible for this moment to propagate transverse to its direction or to locally accumulate in a small volume away from interfaces. We also demonstrate the possibility of orbital moment oscillations induced by lattice potential analogous to spin precession induced by the exchange field of ferromagnets. Thus, in contrast to spin, orbital moments can be controlled by the electron wavevector, without the need for magnetic fields, orbital moment injection or SOC. One can expect rapid relaxation of orbital moment injected into fcc or hcp transition metals, since at least some of the electron hopping is always non-collinear with the orbital moment. This does not contradict the existence of OHE or its importance for orbital moment generation and SHE. Nevertheless, it places stringent constraints on the geometry and crystal structure of devices where atomic orbital moment is generated by OHE in one material, and is injected into another. On the other hand, the interatomic contribution to orbital moment is not constrained by the identified relaxation mechanism, warranting more detailed analyses of different contributions to orbital moment transport.

Recent observations of anomalous current-induced torques in transition metal-based magnetic heterostructures were attributed to orbital moment generation via OHE and its long-range propagation through transition metals such as Pt and Ni~\cite{Hayashi2023-rg,PhysRevB.107.134423}. In the studied geometries, OHE generated orbital moments parallel to the thin-film interfaces, which were assumed to become injected across the interfaces and diffuse through a significant thickness of ferromagnetic layers. This is precisely the transverse geometry that according to our analysis does not allow for atomic orbital moment transport. It is possible that the observed effects resulted entirely from the interatomic orbital moment contribution. We also propose two alternative explanations for the anomalous observations. First, orbitally-selective hopping in transition metals can stabilize an orbital liquid $-$ an orbitally correlated state of electrons that cannot be described in single-particle terms and can exist in both magnetic and non-magnetic materials~\cite{Urazhdin_F}. Orbital correlations are ferromagnetic in the direction normal to the orbital moment, and may mediate long-range orbital torques by analogy to the spin exchange stiffness in ferromagnets. This many-particle mechanism is consistent with the interpretation in terms of orbital moments, but cannot be described by the single-particle picture.

One of the central experimental observations attributed to the orbital moment injection is the variation of STT effects at large ferromagnet thicknesses~\cite{Hayashi2023-rg,PhysRevB.107.134423} inconsistent with the usual STT whose length scale of a few atomic spacings is determined by the rapid dephasing of precession of spin transverse to the magnetization~\cite{Slonczewski1996,Ralph2007SpinTT}. In contrast, the collinear to the magnetization spin  propagates over a much larger longitudinal spin diffusion length. The observed long-range effects may thus be associated with the longitudinal spin transfer whose role in current-induced phenomena remains poorly understood~\cite{PhysRevLett.119.257201,PhysRevLett.126.037203}. These possibilities warrant more detailed studies of the symmetries underlying spin and orbital transport in solids.

This work was supported by the NSF award ECCS-2005786.

\bibliography{OHE}
\bibliographystyle{apsrev4-2}
\end{document}